\documentstyle[aps,preprint]{revtex}

\begin{document}
\preprint{LA-UR-94-534}

\title{
The Influence of Magnetic Imperfections on the Low \\Temperature Properties
of D-wave Superconductors}
\author{A. V. Balatsky}
\address{
 Los Alamos National Laboratory, Los Alamos, NM 87545 and\\ Landau
Institute for Theoretical Physics, Moscow, Russia
}
\author{P. Monthoux}
\address{
Institute for Theoretical Physics, University of California, Santa Barbara,
CA 93106-9530
}
\author{D. Pines}
\address{
Department of Physics, University of Illinois,\\ 1110 West Green Street,
Urbana, IL 61801-3080 }
\date{\today}

\maketitle
\begin{abstract}
We consider the influence of planar ``magnetic" imperfections which destroy the
local magnetic order, such as Zn impurities or $Cu^{2+}$ vacancies, on the low
temperature properties of the cuprate superconductors.  In the unitary limit,
at
low temperatures, for a $d_{x^2-y^2}$ pairing state such imperfections produce
low energy quasiparticles with an anistropic spectrum in the vicinity of the
nodes. We find that for the $La_{2-x}Sr_xCuO_4$ system, one is in the
{\em quasi-one-dimensional} regime of quasiparticle scattering, discussed
recently by Altshuler, Balatsky, and Rosengren, for impurity concentrations in
excess of $\sim 0.16\%$ whereas YBCO$_7$ appears likely to be in the
true 2D scattering regime for Zn concentrations less than $1.6\%$.  We show the
neutron scattering results of Mason et al. \cite{Aeppli} on
$La_{1.86}Sr_{0.14}CuO_4$ provide strong evidence for ``dirty d-wave"
superconductivity in their samples.  We obtain simple expressions for the
dynamic spin susceptibility and $^{63}Cu$ spin-lattice relaxation time,
$^{63}T_1$, in the superconducting state.


\end{abstract}


A number of recent experiments on the cuprate superconductors support the
d$_{x^2-y^2}$ pairing state predicted in spin-fluctuation exchange models
\cite{Pines,Ueda} as well as in Hubbard models \cite{Doug} of this strongly
correlated electron system. For this pairing state, impurities act as
pair-breakers, producing a finite lifetime for quasiparticles near the
nodes  in the gap, and a finite density of states  at low energy
\cite{Rice}. As a result the measured low temperature properties of, for
example, YBCO$_7$ \cite{Bonn}, BiSrCaCO \cite{Takigawa}, and
La$_{1.86}$Sr$_{0.14}$CuO$_4$ \cite{Aeppli}, display a remarkable
sensitivity to the presence of impurities.

Recently the role of imperfections in the d-wave superconductor has been
considered by Lee \cite{Lee} and Hirschfeld and Goldenfeld (hereafter
LHG)\cite{LHG}  and subsequently by others \cite{Doug2,MP2}.  Zn
impurities, which are believed to scatter at the  unitary limit
\cite{MP2}, produce a finite quasiparticle scattering rate, $\gamma  = -
Im \Sigma(\omega \to 0)$, and a finite density of states, $N_{LHG}$, at low
energy in superconducting state:

\begin{eqnarray} \gamma_{LHG}  \simeq \Delta_0
\sqrt{\Gamma/\Delta_0} = \Gamma \sqrt{\Delta_0/\Gamma}. \label{eq1a}\\
N_{LHG} (\omega \to 0)/N_0 \simeq \sqrt{\Gamma/\Delta_0} \sim
n_i^{1/2}. \label{eq1b} \end{eqnarray}

Here we use standard notation: $\Delta_0$  is the amplitude of the gap,
$N_0$ is the density of states at the Fermi surface, $n_i$ is the density of
impurities, and  \begin{equation}
\Gamma \simeq n_i {\epsilon_F\over{\pi}}\end{equation} is the impurity induced
scattering rate in the normal state. The low energy quasiparticle scattering
leads to  experimentally observed consequences in the impurity dominated low
temperature region, $T \leq T^* \simeq \gamma_{LHG}\simeq \Delta_0
\sqrt{\Gamma/\Delta_0}$. All these results were obtained in the assumption that
the impurity is well screened and the effective range of the potential is
``zero".

On the other hand Rosengren, Altshuler and one of the authors (hereafter
ABR) \cite{ABR} pointed out the possibility of a new scattering regime
for a finite range impurity potential in the unitary limit, which they
called the {\em quasi-one-dimensional}(hereafter Q1D) regime. In this
regime quasiparticle dispersion along the Fermi surface is negligible
compared to the scattering rate. Attention was called to the possibility of
finite range impurity potentials by Monthoux and Pines \cite{MP2}, who
pointed out that for cuprates with substantial  antiferromagnetic
correlations, any impurity (such as Zn) which destroys the local magnetic
order, will possess a range comparable to the antiferromagnetic correlation
length, $\xi_{AFM} \sim 2-8 a$. Since the   cuprates have a comparatively
short superconducting coherence length, $\xi \sim 20 \AA$, this implies
that for some systems the impurity potential will  have a  range
comparable  with the superconducting scale. For such a nonzero impurity
potential range, $\lambda$, ABR found a new parameter $\alpha =
{v_F\over{\gamma \lambda}} {\Delta_0\over{\epsilon_F}}$ which measures the
ratio of the energy dispersion of a quasiparticle at the momentum cut-off,
$\lambda^{-1}$, to the scattering rate $\gamma$.  For a ``zero"-range
potential, $\alpha >> 1$, and one recovers the standard unitary scattering
results \cite{Lee,LHG,Doug2}, while for $\alpha \leq 1$, one is in the
quasi-one dimensional regime, with
\begin{eqnarray} \gamma_{ABR}
\simeq \Gamma {\pi^2\over{8}} {\lambda\over{a}} \sim T^*_{ABR}. \label{eq3a}
\end{eqnarray}
As a result, both the temperature $T^*_{ABR} \sim
\gamma_{ABR}$, which defines the impurity-dominated low temperature regime,
and the impurity induced density of states, $N_{ABR}$, vary with the
impurity concentration, rather than with $\sqrt{n_i}$, as in LHG theory.
Specifically, one has: \begin{eqnarray} N_{ABR}/N_0 \sim \gamma/\Delta_0
\sim {\pi\over{8}}(\Gamma/\Delta_0)(\lambda/a) \sim n_i. \label{eq3b}
\end{eqnarray}


In the present communication we examine the applicability of the two
approaches to specific experimental situations. For a small concentration
of Zn impurities  we conclude that YBCO$_7$ is likely an LHG dirty d-wave
superconductor, while a natural explanation of the neutron scattering
results by Mason et.al. \cite{Aeppli} on La$_{1.86}$Sr$_{0.14}$CuO$_4$ in the
superconducting state may be found by applying  ABR theory to the comparatively
large single crystal used in their experiment.



To decide about applicability of LHG (2D)  vs ABR (Q1D) regime  for
impurity scattering in different cuprate systems  we consider the ratio
\begin{equation} {\gamma_{ABR}\over{\gamma_{LHG} }} \simeq \left(
{\Gamma\over{\Delta_0}}\right)^{1/2} \ {\pi^2\over{16}} \
{\xi_{AFM}\over{a}} \end{equation} where we have assumed that $\lambda \sim
\xi_{AFM}/2$, so that near $T_c$, $\lambda \simeq a$ for YBCO$_7$ and
$\lambda = 4a$ for La$_{1.86}$Sr$_{0.14}$CuO$_4$.
 Without calculation it is obvious that in the limit $n_i \to 0$ one is
always in the 2D regime. On introducing  the impurity concentration, in
units of 1\%,  $\tilde{n}_i$, we find specifically: \begin{equation} \Gamma =
10^{-2} \tilde{n}_i {\epsilon_F\over{\pi}} \simeq 3 \tilde{n}_i  \  \ \ \ meV
\label{G} \end{equation} With $\Delta_0 \simeq 12 meV$ for
La$_{1.86}$Sr$_{0.14}$CuO$_4$  (assuming $\Delta_0 = 4 T_c$) and $\Delta_0
\simeq 25 meV$ for YBCO$_7$ we find : \begin{eqnarray}
{\gamma_{ABR}\over{\gamma_{LHG} }} |_{YBCO_7} \simeq 0.4  \
\tilde{n}_i^{1\over{2}} \nonumber\\ {\gamma_{ABR}\over{\gamma_{LHG} }}
|_{La_{0.85}Sr_{0.15}CO_{4} } \simeq 2.5  \ \ \tilde{n}_i^{1\over {2}}
\end{eqnarray} for $\xi_{AFM} \sim 2 (8) a$  for YBCO$_7$
(La$_{1.86}$Sr$_{0.14}$CuO$_4$ ). Hence any concentration of imperfections
which change local magnetic order in excess of $.16   ~\%$, will put
La$_{1.86}$Sr$_{0.14}$CuO$_4$ in the ABR column. On the other hand, for
YBCO$_7$, if $\lambda \sim \xi_{AFM} \sim 2a$ the concentration required to
produce the ABR regime is $\sim 1.6\%$. We caution the reader that these
estimates are based on numerical factors  which are subject to change.

The main reason for such a difference in the concentration of imperfections
required to induce the ABR regime is  that $\xi_{AFM}$ for
La$_{1.86}$Sr$_{0.14}$CuO$_4$ is much larger than for  YBCO$_7$, while the
gap amplitude $\Delta_0$ is smaller.  The parameter $\alpha =
{v_F\over{\gamma \lambda}} {\Delta_0\over{\epsilon_F}}$ is thus much
smaller for  La$_{1.86}$Sr$_{0.14}$CuO$_4$, and it is this which
generates the ABR regime at a significantly smaller concentration of
imperfections for this material.

We now turn  to an analysis of the results  for the inelastic neutron
scattering experiments of Mason et. al. \cite{Aeppli} on
La$_{1.86}$Sr$_{0.14}$CuO$_4$, who found, in the normal state, four
peaks at positions ${\bf Q}_{\delta} = (\pi,\pi) \pm \delta\pi,0);
(\pi,\pi) \pm \delta(0,\pi)$, where $\delta = 0.245$ in units of
$a^{-1}$.  In the superconducting state, Mason et.al. found, quite
unexpectedly, that the intensity of each of the four peaks diminished
only slightly: at $4.2K$ it was only some $60 \%$ of its value just above
$T_c$.
They also found that the intensity of scattering is suppressed ``isotropically"
in the sense that for the momentum transfer ${\bf Q}_{\gamma} = (\pi,\pi) \pm
\delta/2(\pi,-\pi)$ the intensity was suppressed by the same relative amount as
for ${\bf Q}_{\delta}$.  We now show that this is just the behavior
expected for the ABR dirty d-wave superconductor.

Our basic assumption is that at low temperatures the scattering at both ${\bf
Q}_{\delta}$ and ${\bf Q}_{\gamma}$, wavevectors which are separated by
$\delta_q \simeq 0.35(\pi/a)$, is dominated by contribution from the nodal
quasiparticles. Mason et. al. \cite{Aeppli} assume that the Fermi surface for
La$_{1.86}$Sr$_{0.14}$CuO$_4$ is such that node-node scattering  occurs at
momentum transfer ${\bf Q}_{\gamma} = (\pi,\pi) \pm \delta/2(\pi,-\pi)$.
 From Eq~(\ref{eq3a}) and Eq~(\ref{G}) we see that the impurity-induced level
broadening of the nodal quasiparticle states is given by

\addtocounter{equation}{1}
$$\gamma_{ABR} \simeq 3 \left({\lambda \over a}\right)
\tilde{n}_i \simeq 12 \tilde{n}_i \  \ meV \eqno{(\theequation a)}\label{9a}$$

which produces a finite density of states within the node,

$$(N_{ABR}/N_o) \sim {\gamma_{ABR} \over
\Delta_o} \sim \tilde{n}_i \eqno{(\theequation b)}$$.

The characteristic size in the
momentum space of this impurity induced region is on the scale of:

\begin{equation} \delta q \sim {p_F\gamma \over \Delta_0} \sim \pi
\tilde{n}_i/a
\nonumber
\end{equation}

\noindent Hence for an impurity concentration in excess of $.3\%$,
quasiparticle
states both at ${\bf Q}_{\gamma}$ and ${\bf Q}_{\delta}$ are equally affected
by
impurity induced broadening. This effect  naturally explains the ``isotropic"
suppression of neutron scattering intensity \cite{Aeppli}. We expect the
quasiparticle broadening to be anisotropic due to the anisotropy of the
spectrum,
with the greatest broadening occurring within a region $\delta q$ along the
Fermi
surface, rather than perpendicular to it.

At low energy transfer, the scattering intensity at each peak in the normal
state may be expressed in the form introduced by Millis, Monien and Pines
\cite{MMP}: \begin{equation} \chi"_{\bf Q_{\delta}} \left(
{\omega\over{\omega_{SF}}}\right) \equiv \left( \chi_{{\bf Q}_{\delta}}/
\tilde{\chi}_{{\bf Q}_{\delta}}\right)^2 \tilde{\chi}"_{{\bf
Q}_{\delta}}(\omega) \label{a} \end{equation} where $\chi_{{\bf
Q}_{\delta}}$ is the static spin susceptibility, $\omega_{SF}$ is the
energy of the low frequency relaxational mode associated with the strong
antiferromagnetic correlations, \begin{equation} \omega_{SF} =
{\tilde{\Gamma}_{{\bf Q}_{\delta}} \tilde{\chi}_{{\bf
Q}_{\delta}}\over{\chi_{{\bf Q}_{\delta}}}} \end{equation} and
$\tilde{\chi}_{{\bf Q}_{\delta}}, \tilde{\Gamma}_{{\bf Q}_{\delta}}$ are
the static irreducible unenhanced spin susceptibility and characteristic
energy respectively, defined by the  imaginary part of the irreducible
dynamic susceptibility:  \begin{equation} \tilde{\chi}"_{{\bf
Q}_{\delta}}(\omega) = \tilde{\chi}_{{\bf Q}_{\delta}}
\omega/\tilde{\Gamma}_{{\bf Q}_{\delta}}  \end{equation}
Mason et.  al. \cite{pc} find that $\chi_{{\bf
Q}_{\delta}}\simeq 200 \ \ states/eV$ and $\omega_{SF} \simeq 5
\ \ meV$ at $T = 35K$, just above the superconducting transition
temperature.  On the assumption that  ${\bf Q}_{\delta}$ represents  a
momentum transfer
relevant in the impurity dominated region, we can write in the
superconducting state:
\begin{equation}
\chi"_{s{\bf Q}_{\delta}}(\omega) = \left(\chi_{{\bf
Q}_{\delta}}/\tilde{\chi}_{{\bf Q}_{\delta}}\right)^2
\tilde{\chi}"_{ABR}({\bf Q}_{\delta}, \omega)
\label{d}
\end{equation}
where $\chi"_{ABR}({\bf Q}_{\delta}, \omega)
$ is the contribution to the dynamical susceptibility  from the impurity
broadened quasiparticles,
\begin{eqnarray}
\tilde{\chi}"_{ABR}({\bf Q}_{\delta}, \omega) = (\tilde{\chi}_{{\bf
Q}_{\delta}}/\tilde{\Gamma}_{{\bf Q}_{\delta}})(N_{ABR}(0)/N_0)~ \omega
\simeq \left( \gamma /\Delta_0\right) (\tilde{\chi}_{{\bf
Q}_{\delta}}/\tilde{\Gamma}_{{\bf Q}_{\delta}})~\omega
\label{e}
\end{eqnarray}
In writing Eq~(\ref{d}) we have made the assumption that the
antiferromagnetic correlations in the superconducting state are little
changed from their value near $T_c$. This ansatz is consistent with the
fits to the  NMR  experiments  in the superconducting state \cite{TPL} and
with the microscopic calculations of Monthoux and Scalapino and Pao and Bickers
\cite{MS}.

On comparing Eq~(\ref{e}) and Eq~(\ref{a}) we find that the results of
Mason et.al. can be explained, provided the impurity-induced density of
states is reduced by some $77\%$ from its normal state value, \begin{equation}
{N_{ABR}\over{N_0}} = {\gamma\over{\Delta_0}} = 0.77 \end{equation} Such a
value appears quite reasonable. Taking $\Delta_0 \sim 12  \ \ meV$, we find
$\gamma \simeq 9 \ \ meV$. This is a value of the scattering rate which,
according to Eq (9a), is generated  with only
a $0.6\%$ level of magnetic imperfections which scatter at the unitary limit.
Given the fact that the neutron scattering experiments are done on large
samples
this estimate appears physically reasonable.

We consider briefly the influence of impurities on the Knight shift
and $^{63}Cu$ spin-lattice relaxation rate. For the Knight shift, taking Fermi
liquid corrections into account, one has, below $T^*$, \begin{eqnarray}
{^{63}K^s(T)\over{^{63}K^s(T_c)}} = (1 - F^a_0)~{N_{imp}(0)\over{N_0}}
\end{eqnarray} where $F^a_0$ is the Fermi liquid parameter ($\sim 1/2$ for
YBCO$_7$, smaller for spin gap materials) at $T_c$. This result is applicable
in either the ABR or LHG limit.  The spin-lattice relaxation rate in the normal
state is: \begin{eqnarray} ^{63}(T_1T)^{-1} = \sum_{{\bf q}}
^{\alpha}F_{{\bf q}}
{\chi"({\bf q}, \omega)\over{\omega}} \simeq F_{\alpha} {\chi_{{\bf
Q}}\over{\xi^{2} \omega_{SF}}} \label{T1} \end{eqnarray} for long correlation
lengths, $\xi$; here $F^{\alpha}_q$ and $F_{\alpha}$ are the $^{63}Cu$ form
factor and corresponding constant which depend on the
orientation of the applied magnetic field. In the superconducting state, below
$T^*$, on substituting Eq.(\ref{e}) into  Eq. (\ref{T1}) and carrying out the
momentum sums, we obtain, \begin{eqnarray} {^{63}(T_1T)^{-1}_{imp} \over
(^{63}T_1(T_c)T_c)^{-1}} = \left({N_{imp}(0)\over{N_0}}\right)
\label{T1imp} \end{eqnarray}
These results, which are valid whether one is in the LHG or
ABR regime, appear capable of providing a quantitative account of the results
of Ishida et al. \cite{Bonn} for the influence of Zn impurities on the
$^{63}Cu$ Knight shift in $YBa_2Cu_3O_7$, and of Ishida et al. \cite{Takigawa}
and Takigawa \cite{Takigawa} for the low temperature behavior of both the
$^{63}Cu$ Knight shift and spin lattice relaxation rate in BSrCCO.

\underline{Conclusion} We compared the isotropic unitary scattering regime
(LHG) with the strongly anisotropic scattering in ABR regime, and show that
${\gamma_{ABR}\over{\gamma_{LHG} }} \sim  \sqrt{N_{imp}}$. At small
impurity concentration the scattering is always in the LHG regime. This is
the limit realized in pure and lightly Zn doped YBCO$_7$ . For the
La-based materials we find that at $n_i \geq 0.16\%$ one is always in
the ABR scattering regime. Strong nodal quasiparticle level broadening,
characteristic of ABR regime, occurs in a substantial part of the momentum
space $\delta q/p_F \simeq \gamma_{ABR}/\Delta_0$. This provides a natural
explanation of the ``isotropic" suppression of neutron scattering intensity
in La$_{1.86}$Sr$_{0.14}$CuO$_4$ below $T_c$. We find that our inferred
quasiparticle broadening, $\gamma \simeq 9 \ \ meV$, will require about
$0.75\%$ of imperfections which destroy the local magnetic order. Such
concentrations of imperfections which destroy the local magnetic order appear
plausible for the large samples required to carry out neutron experiments.

\underline{Acknowledgements} We are grateful  to D. Bonn, W. Hardy, P.
Hirschfeld and  P. Littlewood  for useful discussions.  Part of the work
presented here was performed at the Aspen Center for Physics, and part at
Los Alamos under the auspices of the program on strongly correlated
electron systems, whose support is acknowledged. This work was
also supported in part by a J.R. Oppenheimer fellowship (AVB), by NSF
Grant PHY89-04035 (PM), and by NSF Grant DMR91-20000 (DP).


\begin{references}




\bibitem{Pines}
P.\ Monthoux, A.\ V.\ Balatsky and D.\ Pines, Phys.\ Rev.\ Lett.\ {\bf 67},
3448 (1991); P.\ Monthoux and D.\ Pines, Phys.\ Rev.\  {\bf B 47}, 6069
(1993); For a recent review see D. Pines, Physica B, in the press.



\bibitem{Ueda}
T.\ Moriya, Y.\ Takahashi and K.\ Ueda, Physica {\bf C 185-189}, 114,
(1991); K.\ Ueda, T.\ Moriya and Y.\ Takahashi, in {\em Electronic
Properties and Mechanisms of High-$T_c$ Superconductors}, Ed.\ Oguchi,
(North Holland, 1992), p.\ 145.

\bibitem{Doug}
G.\ Kotliar and J.\ Liu, Phys.\ Rev.\ {\bf B 37}, 5142 (1988); S.\ R.\ White,
D.\ J.\ Scalapino, R.\ Sugar, N.\ E.\ Bickers and R.\ Scalettar, Phys.\ Rev.\
{\bf B 39}, 839 (1989).

\bibitem{Rice}
K.\ Ueda and T.\ M.\ Rice, {\em Theory of Heavy Fermions and Valence
Fluctuations}, Ed.\ T.\ Kasuya and T.\ Saso, (Springer, Berlin, 1985), p.\
267;
L.\ P.\ Gor'kov and P.\ A.\ Kalugin, Pisma ZhETP, {\bf 41}, 208, (1985);
JETP Lett., {\bf 41}, 253 (1985).




\bibitem{Bonn}
D.\ Bonn {\em et al.}, preprint, 1993; K. Ishida et al., Physica C {\bf
185-189}, 1115 (1991).


\bibitem{Takigawa} M. Takigawa, private communication; K. Ishida et. al.,
preprint.

\bibitem{Aeppli}
T.\ Mason {\em et al.}, Phys.\ Rev.\ Lett.\ {\bf 71}, 919 (1993).

\bibitem{Lee}
P.\ Lee, Phys.\ Rev.\ Lett.\ {\bf 71}, 1887, (1993).

\bibitem{LHG}
P.\ J.\ Hirschfeld and N.\ Goldenfeld, Phys.\ Rev.\ {\bf B 48}, 4219 (1993); T.
Hotta, J. Phys. Soc. Jpn. {\bf 62}, 274 (1993).





\bibitem{Doug2}
P.\ J.\ Hirschfeld, W.\ O.\ Putikka and D.\ J.\ Scalapino, preprint, 1993;
unpublished.




\bibitem{MP2}
P.\ Monthoux and D.\ Pines, Phys. Rev {\bf B}, in the  press, (1994).




\bibitem{ABR} A.V. Balatsky, A. Rosengren and B.L. Altshuler, preprint, 1993.


\bibitem{MMP} A. Millis, H. Monien and D. Pines, Phys. Rev. {\bf B 42},
167, (1990).

\bibitem{pc} T. Mason, private communication.


\bibitem{TPL} J. Martindale {\em et al.}, Phys. Rev. {\bf B 47},
1955 (1993); D. Thelen. D. Pines and J.P. Lu, Phys. Rev.\ {\bf B 47}, 9151
(1993).

\bibitem{MS} P. Monthoux and D.J. Scalapino, preprint, 1993; C. Pao and N.
E. Bickers, preprint, 1993.


\end{references}
\end{document}